\documentclass[aps,prl,twocolumn, titlepage,showpacs]{revtex4}
\usepackage{graphicx}
\usepackage{color}
\usepackage{dcolumn}

\usepackage{amsfonts}

\usepackage{bm}

\begin{document}
\title{Structures and Diffusion of Two Dimensional Dusty Plasmas on One Dimensional Periodic Substrates}

\author{Kang Wang$^1$, Wei Li$^1$, Dong Huang$^1$, C. Reichhardt$^2$, C. J. O. Reichhardt$^2$, M. S. Murillo$^3$, and Yan Feng$^1$}

\email{fengyan@suda.edu.cn}

\affiliation{
$^1$ Center for Soft Condensed Matter Physics and Interdisciplinary Research, School of Physical Science and Technology, Soochow University, Suzhou 215006, China\\
$^2$ Theoretical Division, Los Alamos National Laboratory, Los Alamos, New Mexico 87545, USA\\
$^3$ Department of Computational Mathematics, Science and Engineering, Michigan State University, East Lansing, Michigan 48824, USA\\}

\date{\today}

\begin{abstract}

Using numerical simulations, we examine the structure and diffusion of a two-dimensional dusty plasma (2DDP) in the presence of a one-dimensional periodic substrate (1DPS) as a function of increasing substrate strength. Both the pair correlation function perpendicular to the substrate modulation and the mean squared displacement (MSD) of dust particles are calculated. It is found that both the structure and dynamics of 2DDP exhibit strong anisotropic effects, due to the applied 1DPS. As the substrate strength increases from 0, the structure order of dusty plasma along each potential well of 1DPS increases first probably due to the competition between the inter-particle interactions and the particle-substrate interactions, and then decreases gradually, which may be due to the reduced dimensionality and the enhanced fluctuations. The obtained MSD along potential wells of 1DPS clearly shows three processes of diffusion in our studied 2DDP. Between the initial ballistic and finally diffusive motion, there is the intermediate sub-diffusion discovered here, which may result from the substrate-induced distortion of the caging dynamics.

\end{abstract}

\pacs{64.30.-t, 52.27.Gr, 52.27.Lw}\narrowtext

\maketitle

\section{I.~Introduction}

There are a variety of two-dimensional (2D) systems that can be effectively described as a collection of particles interacting with a periodic one-dimensional (1D) substrate. Specific examples of such systems include charged colloids on optical trap arrays~\cite{Chowdhury:1985, Wei:1998, Frey:1999, Bechinger:2001, Zaidouny:2013}, magnetic particles on patterned surfaces~\cite{Hu:1997, Tierno:2012}, vortices in type-II superconductors with 1D periodic pinning arrays~\cite{Martinoli:1978, Levitov:1991, Jaquel:2002, Moll:2014, Guillamon:2014, Le:2016}, vortices in Bose-Einstein condensates on 1D optical arrays~\cite{Reijnders:2004}, and the ordering of molecules and atoms on anisotropic surfaces~\cite{Coppersmith:1982, Vanossi:2013}. In these systems, a variety of distinct commensurate-incommensurate transitions occur along with multiple ordering or disordering transitions as functions of the substrate strength or the ratio of the number of particles to the number of substrate minima, known as the filling factor.

Colloidal particles interacting with optical substrates are a particularly convenient system in which to study such effects since, due to the size scale of the colloids, it is possible to directly access the microscopic details of the particle positions and dynamics~\cite{Chowdhury:1985, Wei:1998, Frey:1999, Bechinger:2001}. Structural and dynamical properties of 2D colloids on an external periodic substrate have been widely investigated, and intriguing phase behaviors and rich dynamics have been observed over the past decades~\cite{Chowdhury:1985, Wei:1998, Reichhardt:2002, Baumgartl:2004, Bohlein1:2012}. Several methods have been used to realize the periodic potential in colloids experiments, for example interfering of laser beams~\cite{Chowdhury:1985, Wei:1998}, periodic pinning arrays~\cite{Mangold:2003}, and imprint or stamping techniques~\cite{Lin:2000}. On a 1DPS, in 1985, Chowdhury et al., firstly observed the intriguing new phenomena, laser-induced freezing in 2D colloids on a one-dimensional periodic substrate (1DPS), provided by two interfering laser beams, with increasing the intensity of the laser~\cite{Chowdhury:1985}. Hereafter, Loudiyi et al. directly observed the laser-induced freezing in colloids experiments~\cite{Loudiyi1:1992} and simulations~\cite{Loudiyi2:1992}, respectively. The reentrant melting, or laser-induced melting, was observed by Wei~\cite{Wei:1998}, with further increasing the intensity of the laser in colloids experiments. On a 1DPS, subdiffusive colloid motion occurs at intermediate time scales~\cite{Herrera-Velarde:2009}. When the channels are so narrow that the colloids cannot pass each other, single-file diffusion (SFD) can occur~\cite{Herrera-Velarde:2007} with a mean-square displacement (MSD) that follows the form $W(t) = 2Ft^{1/2}$ on long time scales, where $F$ is the SFD mobility factor. Other rich dynamics of the colloidal monolayer, such as pinning and depinning~\cite{Reichhardt:2005}, friction dynamics~\cite{Korda:2002}, kinks and anti-kinks~\cite{Bohlein2:2012}, and topological defect dynamics~\cite{Kromer:2012, McDermott:2013} have also been investigated.

On the same size scale as the colloidal system is a dusty plasma, consisting of a partially ionized gas containing highly charged micron-sized dust particles~\cite{Fortov:2005, Morfill:2009, Piel:2010, Bonitz:2010, Merlino:2004, Feng:2008, Thomas:2004}. In the laboratory conditions, these dust particles typically have a negative charge of $Q \approx -10^{4}e$, which can be directly imaged~\cite{Thomas:1994, Chiang:1996, Quinn:2001, Nosenko:2002, Liu:2008, Morfill:2009} with the typical distance of $0.2~{\rm mm}$ between nearest dust particles. By balancing gravity with an electric field, the charged dust particles can be levitated and confined in the plasma sheath, where they self-organize into a single-layer suspension with the negligible out-of-plane motion, called a 2D dusty plasma (2DDP)~\cite{Feng:2011}. Due to the shielding effect of the electrons and ions in the plasma~\cite{Yukawa:1935, Konopka:2000, Zhang:2010}, the in-plane interaction between dust particles is a Yukawa potential, and the interactions are typically so strong that the collection of these dust particles exhbibits the liquid~\cite{Kalman:2004, Chan:2007, Feng:2010, Ott:2011, Feng:2012, Rosenberg:2012, Ott:2015, Haralson:2017} or solid~\cite{Feng:2008, Chu:1994, Thomas:1996, Melzer:1996, Hartmann:2010, Feng1:2010, Feng2:2010, Hartmann:2014} behaviors. Dusty plasmas have been used to study 2D melting~\cite{Quinn:2001}, diffusion~\cite{Liu:2008}, dislocation dynamics~\cite{Durniak:2013}, and fluctuation relations~\cite{Wong:2018}. Langevin dynamical simulations of Yukawa liquids are widely employed to investigate the behaviors of dusty plasmas~\cite{Feng1:2011, Feng:2017, Schweigert:2000, Donko:2010, Melzer:2013}.

A key difference between colloidal systems and dusty plasmas is that the fluid in which the colloids are immersed causes their dynamics to be overdamped, while dusty plasmas are underdamped and can exhibit phenomena such as phonon propagation, shock waves, and other effects~\cite{Morfill:2009}. Although the transport of dusty plasmas on substrates was investigated by~\cite{Li:2009, Li:2010}, and the collective phonon spectra modes were described in~\cite{Li:2018}, most studies of dusty plasmas have been performed without a substrate, so commensuration effects, diffusion, and order-disorder transitions have not yet been explored.

The rest of this paper is organized as follows. In Sec.~II, we will briefly introduce the Langevin dynamical simulations, with external substrates. Then, in Sec.~III, we will discuss our results on the structural and dynamical properties, respectively, from our calculated pair correlation function in the $y$ direction $g(y)$ and mean squared displacement. Finally, we conclude with a brief summary.

\section{II.~Simulation Methods}

We use Langevin dynamical simulations to represent 2D dusty plasmas on 1DPS, where the equation of motion for dust particle $i$ is given by
\begin{equation}\label{LDS}
{m \ddot{\bf r}_{i} = -\nabla \Sigma \phi_{ij} - \nu m\dot{\bf r}_{i} + \xi_i(t) + {\bf F}^{S}_{i}.}
\end{equation}
Here, $-\nabla \Sigma_{j} \phi_{ij}$ is the particle-particle interaction force, with a Yukawa potential, $\phi_{ij} = Q^2\exp(-r_{ij}/\lambda_D)/4 \pi \epsilon_0r_{ij}$ ($Q$ is the charge of each dust particles, $\lambda_D$ is the screening length, and $r_{ij}$ is the distance between dust particles $i$ and $j$), $- \nu m\dot{\bf r}_{i}$ is the frictional drag, and $\xi_i(t)$ is the noise term; the Langevin terms satisfy the fluctuation-dissipation theorem for dust particles in a plasma background with dust-neutral damping. The last term on the right-hand side of Eq.~(\ref{LDS}), ${\bf F}^{S}_{i}$, is the external force exerted by the 1D periodic substrate, which has the form
\begin{equation}\label{1DPS}
{U(x) = U_{0} \cos(2 \pi x/w),}
\end{equation}
giving ${\bf F}^{S}_{i} = - \frac {\partial U(x)}{\partial x} \hat{\bf x} = (2\pi U_{0}/w)\sin(2\pi x/w) \hat{\bf x}$. Here, $U_{0}$ and $w$ are the depth and width of the potential well, respectively. Note that here the depth of the potential well $U_{0}$ can also be regarded as the substrate strength.

Two dimensionless parameters, the coupling parameter $\Gamma = Q^2/(4\pi \epsilon_0 ak_{B}T)$ and the screening parameter $\kappa \equiv a/\lambda_{D}$ can be used to characterize our simulated 2D dusty plasmas. Here, $T$ is the kinetic temperature of the dust particles, $a = (\pi n)^{-1/2}$ is the Wigner-Seitz radius, and $n$ is the areal number density. The inverse of nominal 2D dusty plasma frequency $\omega_{pd}^{-1} = (Q^2/2\pi \epsilon_0 ma^3)^{-1/2}$ is used to normalize the time scale, and the Wigner-Seitz radius $a$ or the lattice constant $b$ is used to normalize the length~\cite{Wang:2018}. We use $k_{B}T$ to normalize the depth of the potential well $U_{0}$. Note, we can also use $E_{0}=Q^2/4\pi \epsilon_0 b$ to normalize the depth of the potential well $U_{0}$, as in~\cite{Li:2018}.

We simulate $N = 1024$ dust particles constrained to lie within a 2D plane of dimensions $61.1a \times 52.9a$ with periodic boundary conditions, and we choose the constant values of $\Gamma = 200$ and $\kappa = 2.0$. Since the simulated size in the $x$ direction is $61.1a \approx 32.07b$, we set the width of the potential well to $w = 1.002b$, corresponding to $32$ full potential wells within the simulation box to satisfy the periodic boundary conditions. We vary the depth of the potential well $U_{0}/k_{B}T$ over the range $U_{0}/k_{B}T = 0$ to $10.50$. The gas damping we choose is $\nu /\omega_{pd} = 0.027$, comparable to the typical dusty plasma experiments~\cite{Feng1:2008}. We integrate Eq.~(\ref{LDS}) for $10^6$ steps using a time step of $0.037\omega_{pd}^{-1}$ using the Yukawa potential that is truncated at a distance beyond a cutoff radius of $24.8a$ as in~\cite{Feng:2013}. Other simulation details are similar to~\cite{Li:2018}. For each simulation run, we begin with a random configuration of dust particles and integrate for $3 \times 10^5$, simulation time steps  at a desired temperature to achieve the final steady condition. We then record the positions and velocities of all $N = 1024$ particles in the next $10^6$ steps for later data analysis.

\section{III.~Results and Discussion}

\subsection{A.~Structures of 2DDP on 1DPS}

\begin{figure}[htb]
	\centering
        	\includegraphics{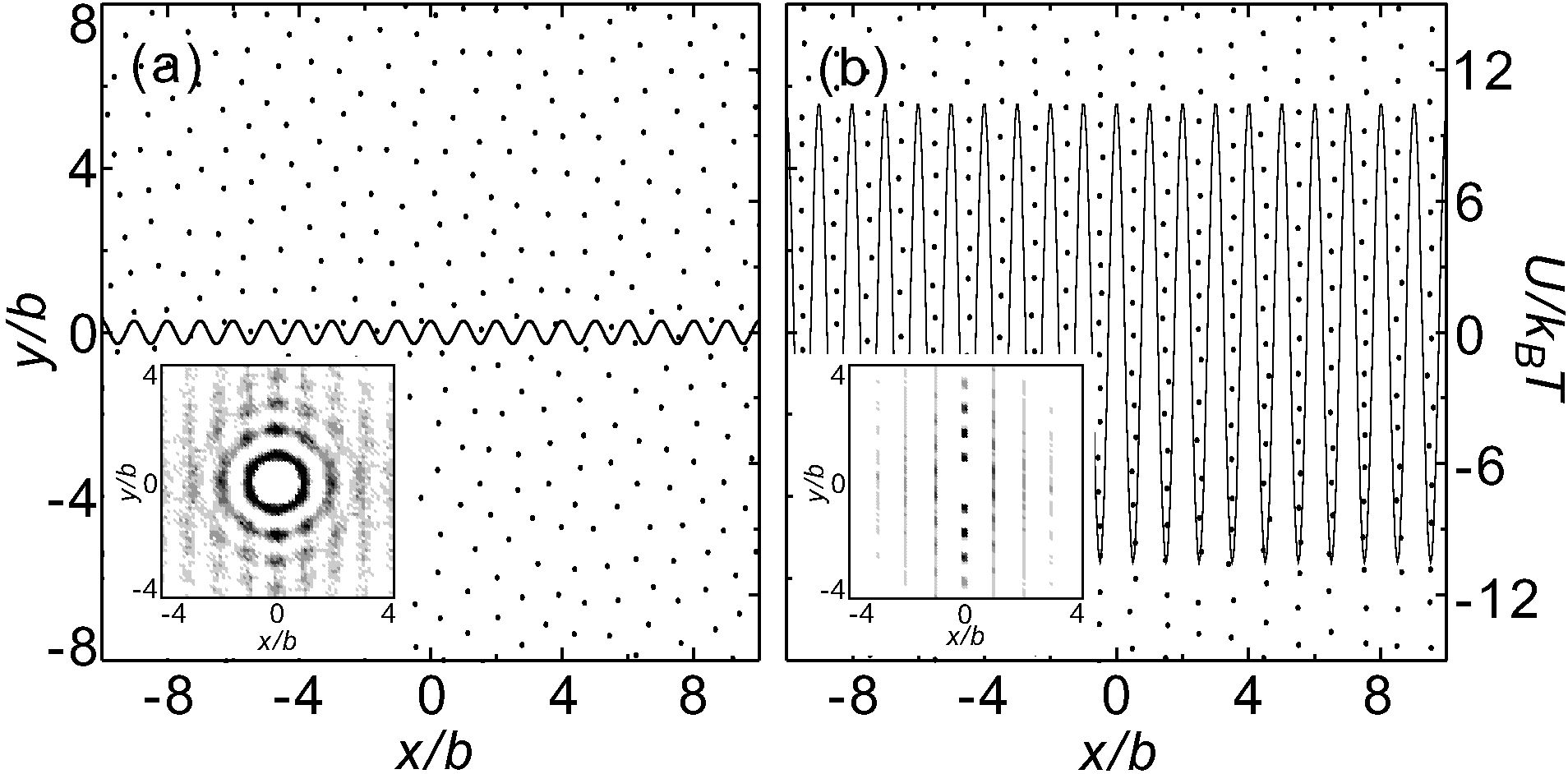}
	\caption{\label{1DPS} Snapshots of the simulated dust particles positions (black dots) within different 1DPS (solid curves). When the depth of 1DPS is small, $U_{0}/k_{B}T = 0.525$ in (a), the distribution of dust particles is random with the typical liquid state. When the depth of 1DPS is larger, $U_{0}/k_{B}T = 10.50$ in (b), the dust particles are confined in the different columns provided by the 1DPS, forming several 1D or quasi-1D chains. For each panel, the inset on the bottom left corner is the 2D distribution function~\cite{Loudiyi1:1992} $g(x,y)$. Note, only $\approx 36\%$ of the total simulated region is plotted here.
    }
\end{figure}

Firstly, we focus on the structural properties of 2DDP on 1DPS, with different depths of the potential wells of the applied substrate. In Fig.~1(a) we illustrate the positions of the particles along with the shape of the 1DPS for a substrate with a weak substrate strength of $U_0/k_BT=0.525$. The particles adopt a liquid ordering, as indicated by the 2D distribution function $g(x,y)$ calculated as described in~\cite{Loudiyi1:1992}, and shown in the lower left corner of the figure. When the substrate is stronger, the particles are no longer isotropically distributed and form quasi-1D rows aligned with the substrate minima, as shown in Fig.~1(b) for $U_0/k_BT=10.5$.

\begin{figure}[htb]
	\centering
        	\includegraphics{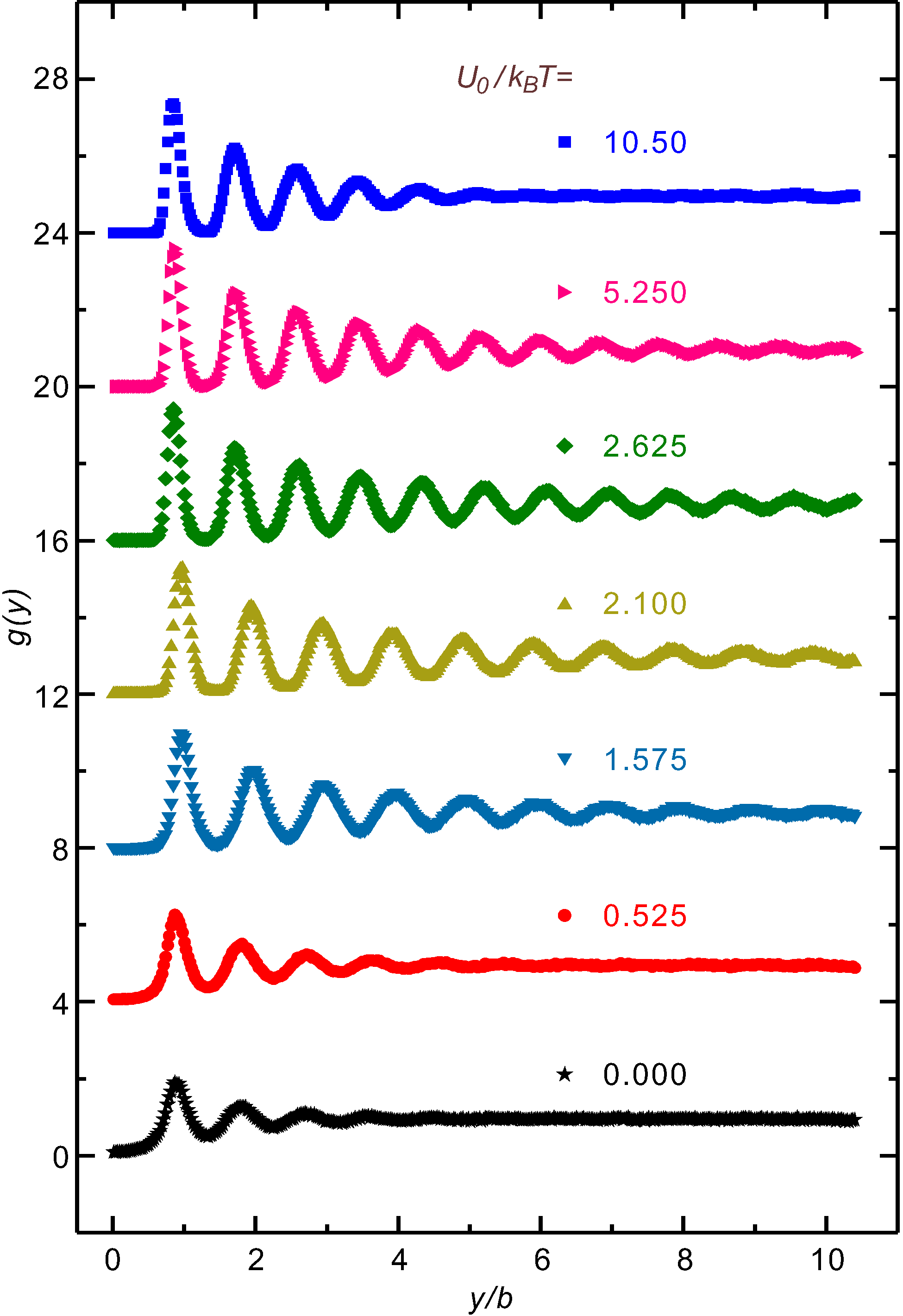}
	\caption{\label{PCFY1}(Color online). Averaged pair correlation functions $g(y)$ in the $y$ direction, parallel to the potential wells of 1DPS, for various depths of 1DPS, $U_0/k_BT=0$, 0.525, 1.575, 2.1, 2.625, 5.25, and 10.5, from bottom to top. The curves have been offset vertically for clarity. As the depth of the 1DPS $U_0$ increases gradually from zero, the number of peaks in $g(y)$ first increases, and then gradually decreases as the structure transitions from the initial disordered liquid state to an intermediate modulated order state, followed by a reentrant melting transition into a modulated disordered state.
	}
\end{figure}

Due to the anisotropy of our studied 2DDP caused by the 1DPS, we calculate the averaged pair correlation function only in the $y$ direction (PCFY) $g(y)$, along the potential well of the 1DPS. We divide the system into 32 columns corresponding to the number of substrate minima, compute the value of $g(y)$ for each column, and finally average $g(y)$ over all columns. This is because the positions of the dust particles in neighboring columns are not correlated with each other when the substrate is sufficiently strong~\cite{Li:2018}.

In Fig.~2, we plot various $g(y)$ for different strength of 1DPS ranging from $U_0/k_BT=0$ to 10.50. In the absence of a substrate, when $U_{0}/k_{B}T = 0$, $g(y)$ decays quickly, indicating that the system is in a disordered liquid state, as expected for our simulation Yukawa conditions of $\Gamma = 200$ and $\kappa = 2$. As $U_{0}/k_{B}T$ increases from 0 to 2.625, $g(y)$ decays much more slowly with $y$ and the number of peaks in $g(y)$ increases substantially. This gradual emergence of a modulated ordered state is probably due to the competition between the particle-particle interactions and the particle-substrate interactions. While, when $U_{0}/k_{B}T$ increases 2.625 to 10.50, the decay of $g(y)$ becomes more rapid and reentrant melting occurs into a modulated disordered state. We speculate this reentrant melting probably results from the increasing constraint of particle motion in the $x$ direction by the substrate, which would result in a reduced effective dimensionality of the system and simultaneously enhance the effect of fluctuations~\cite{Wei:1998}.

\begin{figure}[htb]
	\centering
        	\includegraphics{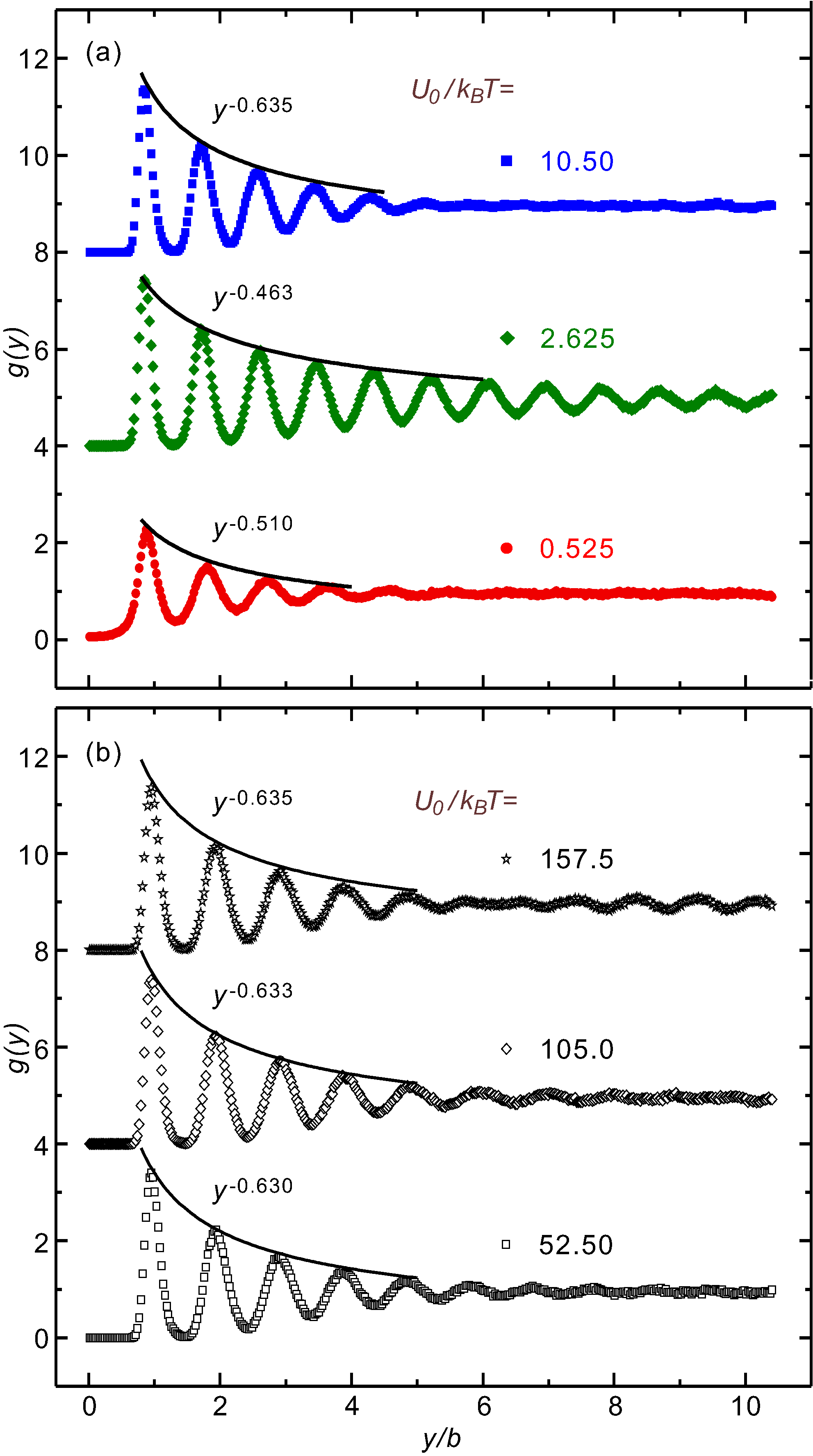}
	\caption{\label{PCFY2}(Color online) The averaged pair correlation $g(y)$ (symbols) along with fits (solid lines) to $g(y) - 1 \sim y^{-\gamma}$ for the peaks in $g(y)$. The curves have been offset vertically for clarity. (a) Substrates with $U_{0}/k_{B}T = 0.525$, 2.625 and 10.50, from bottom to top. For the intermediate substrates with $U_0/k_BT=2.625$, we find that in this modulated ordered state $g(y)$ decays the most slowly with the smallest exponent of $\gamma = 0.463$. (b) Substrates with $U_{0}/k_{B}T =52.5$, 105.0, and 157.5, from bottom to top. For strong substrates, the exponents are nearly constant, $\gamma \approx 0.63$, so that further increase in the strength of substrates has little effect on the structure in the $y$ direction.
	}
\end{figure}

The phase transition results we discover above have not been reported in dusty plasmas before, either in simulations or in experiments. However, similar results of the laser-induced freezing and laser-induced melting~\cite{Wei:1998, Bechinger:2001} have been observed in colloids on 1DPS provided by the interference of two laser beams.

To quantify the structural changes, we also fit the decay of the heights of the peaks in $g(y)$ of 2DDP on different depths of 1DPS to the power law $g(y) - 1 \propto y^{- \gamma}$, as solid lines shown in Fig.~3. For both the weak and strong of 1DPS, or the value of $U_{0}/k_{B}T$ is small or large, the $g(y)$ decays much faster and $\gamma$ is pretty large. However, at the intermediate level of 1DPS, such as $U_{0}/k_{B}T = 2.625$, $g(y)$ decays much more slowly, with a much smaller exponent of $\gamma = 0.46$, corresponding to the modulated ordered state. In addition, we also fit the $g(y)$ with much stronger strength of 1DPS, more than one order of magnitude as shown in Fig.~3(b). However, the corresponding $g(y)$ with much stronger depths of 1DPS decays with a nearly constant exponent of $\gamma \approx 0.63$, indicating that the depth of 1DPS has little effect on the structure in the $y$ direction for this modulated disorder state.

\begin{figure}[htb]
	\centering
        	\includegraphics{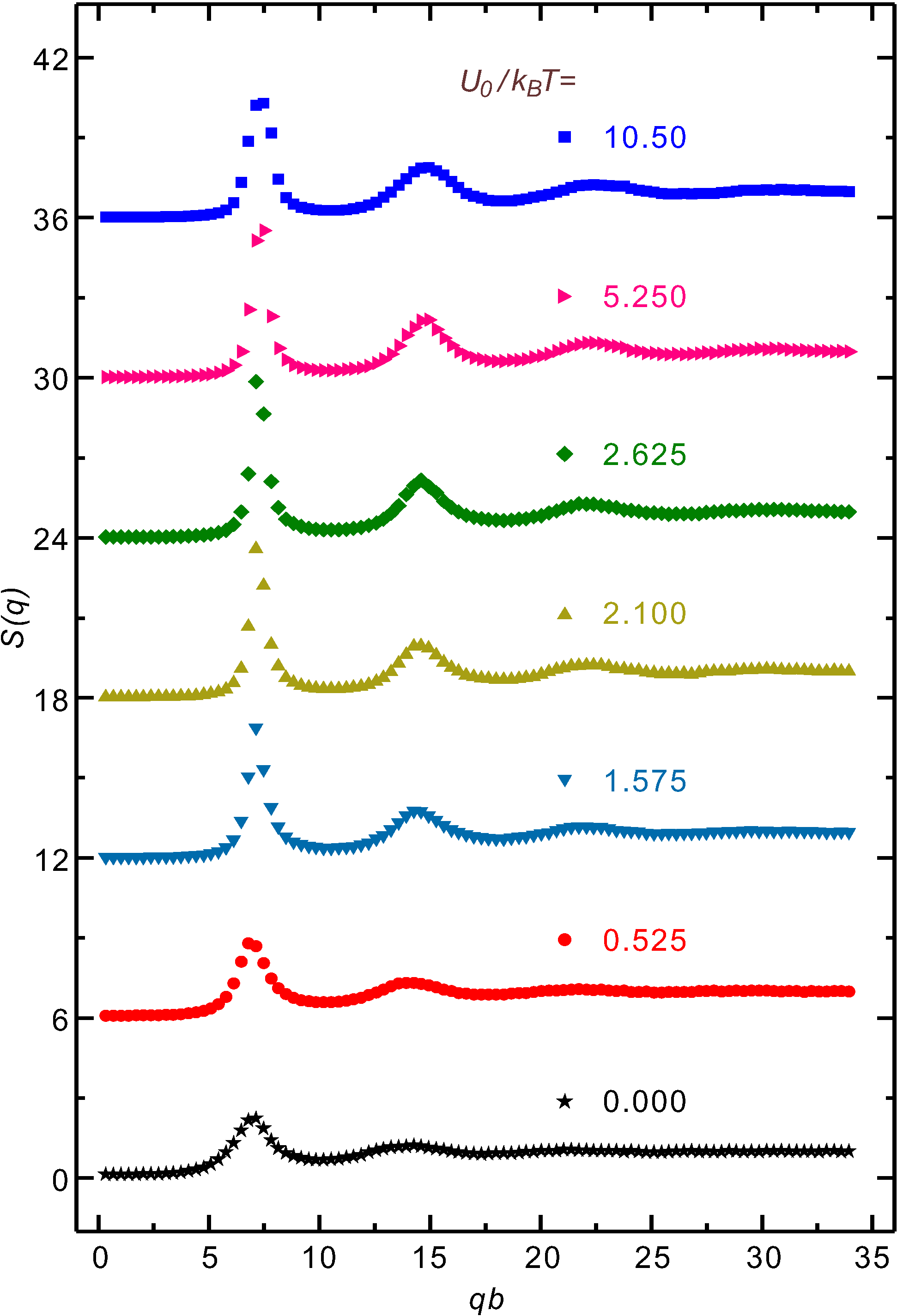}
	\caption{\label{S(q)}(Color online)  The averaged structure factor $S(q)$ in the $y$ direction, along the potential wells of 1DPS, for $U_0/k_BT=10.5$ (squares), 5.25 (right triangles), 2.625 (diamonds), 2.1 (up triangles), 1.575 (down triangles), 0.525 (circles), and 0 (stars). We find the sharpest peaks for intermediate values of $U_0/k_BT$ ranging from 2.1 to 5.25, where the particles develop the strongest 1D ordering along the substrate troughs.
    }
\end{figure}

We also calculate the structure factor $S(q)$ in the $y$ direction, along the potential wells of 1DPS, for the different substrate strength, as shown in Fig.~4. The sharpest peaks, indicating the largest amount of 1D ordering, appear for intermediate substrate strengths of $2.1 \leq U_0/k_BT \leq 5.25$. When the substrate is weaker, the particles are not fully confined to the 1D channels, so the 1D ordering is reduced. While when the substrate is strong, it overwhelms the particle-particle interaction energy and melts the 1D solid in each channel, so that the system is instead composed of a series of 1D liquids.

\begin{figure}[htb]
	\centering
        	\includegraphics{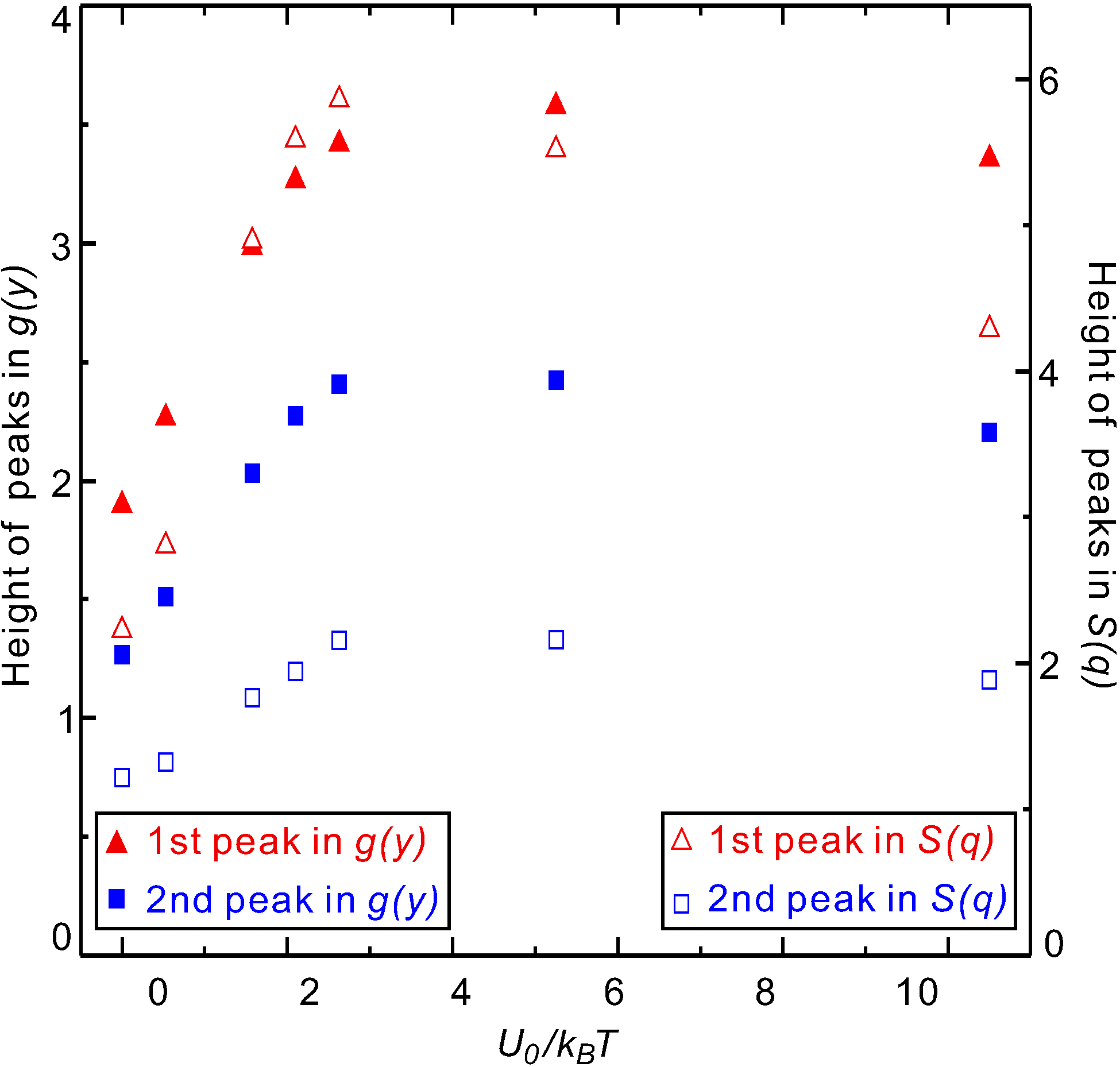}
	\caption{\label{Peaks}(Color online)  The heights of the first and second peaks in $g(y)$ and $S(q)$ for different depths of a 1DPS. From this figure, we can clearly observe the values of peaks firstly increase, corresponding to the more ordered state, and then decrease gradually, corresponding to the reentrant melting into a modulated disorder state.
    }
\end{figure}

Here, we also plot the heights of the first and second peaks in $g(r)$ and $S(q)$ as functions of the depth of a 1DPS $U_{0}/k_{B}T$, as shown in Fig.~5. It is clearly to find that the values of the first and second peaks all firstly increase and then decrease gradually, as the substrate strength increases from 0. These results clearly indicate that the structure changes from the initial disorder liquid state to the intermediate modulated order state, and then a reentrant melting into a modulated disorder state occurs, which is consistent with our previous analysis above.

\subsection{B.~Diffusion of 2DDP on 1DPS}

To study the diffusion of 2DDP on 1DPS, we calculate the mean-square displacement (MSD), defined as

\begin{equation}\label{MSD}
{MSD = \langle |{\bf r}_{i}(t) - {\bf r}_{i}(0)|^{2} \rangle = 4Dt^{\alpha (t)},}
\end{equation}
where ${\bf r}_{i}(t)$ is the position of the ith particle at time $t$, $\langle ~ \rangle$ denotes the ensemble average and $D$ is the diffusion coefficient.

\begin{figure*}[htb]
	\centering
        	\includegraphics{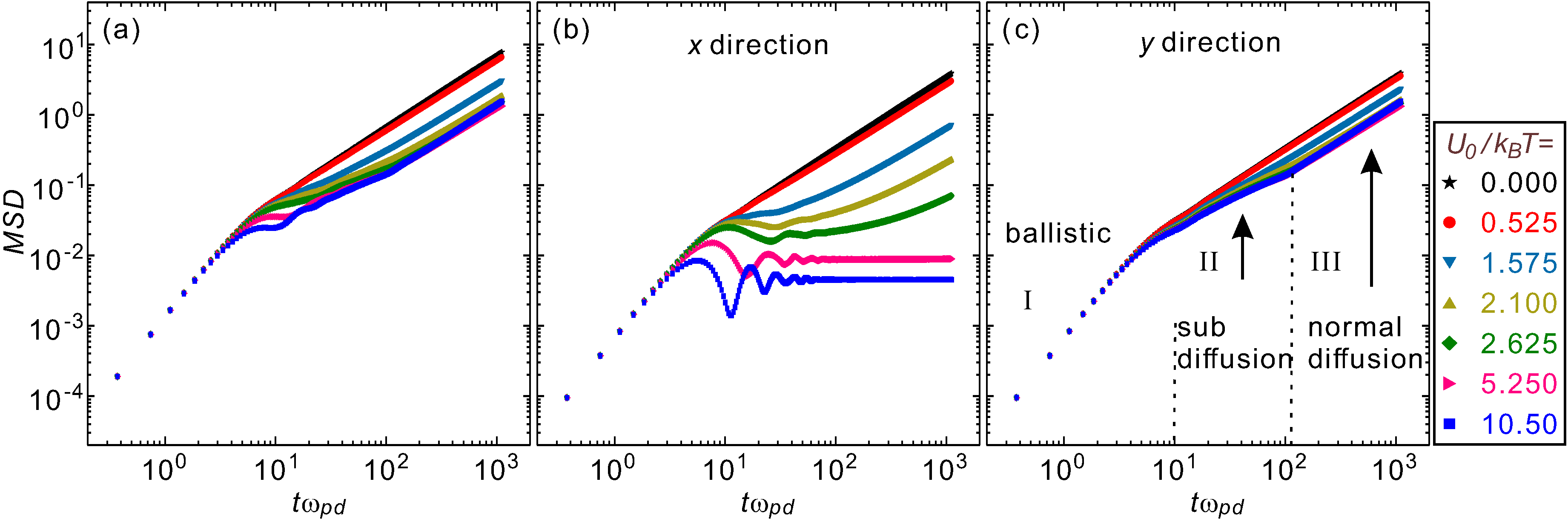}
	\caption{\label{MSD}(Color online) Mean squared displacement (MSD) of different directions of 2D dusty plasmas for different depths of 1DPS. (a) MSD calculated from the motion in both the $x$ and $y$ directions. (b) MSD calculated from the motion in only the $x$ direction (XMSD), perpendicular to the potential wells. (c) MSD calculated from the motion in only the $y$ direction (YMSD), along the potential wells. In the $x$ direction, with stronger strength of 1DPS, the diffusion is completely constraint. In the $y$ direction, the diffusion is suppressed gradually by the 1DPS. When the strength of 1DPS is slightly larger enough, $U_{0}/k_{B}T \ge 2.625$, all YMSD curves overlap together, suggesting YMSD will not be substantially changed as the strength 1DPS increases more.
	}
\end{figure*}

Figure~6 shows the calculated MSD of 2D dusty plasmas in different directions for the different strength of 1DPS. In Fig.~6(a), the MSD computed from the combination of the $x$ and $y$ directions decreases in magnitude as $U_0/k_BT$ increases. For the $x$ direction MSD (XMSD), as in Fig.~6(b), the diffusion becomes completely constant when the 1DPS is strong enough, $U_0/k_BT \ge 5$. In this regime, the diffusion is almost completely contributed by motion in the $y$ direction at longer times, since there is no confinement along potential wells. As shown in Fig.~6(c), the $y$ direction MSD (YMSD) is gradually suppressed as $U_0/k_BT$ increases. When $U_{0}/k_{B}T \ge 2.625$, all of the MSD and YMSD curves nearly overlap, indicating that, once the substrate is sufficiently deep, both the MSD and the YMSD do not vary substantially any more, i.e., the diffusion behavior along the potential wells does not change much while $U_{0}/k_{B}T \ge 2.625$.

\begin{figure}[htb]
	\centering
        	\includegraphics{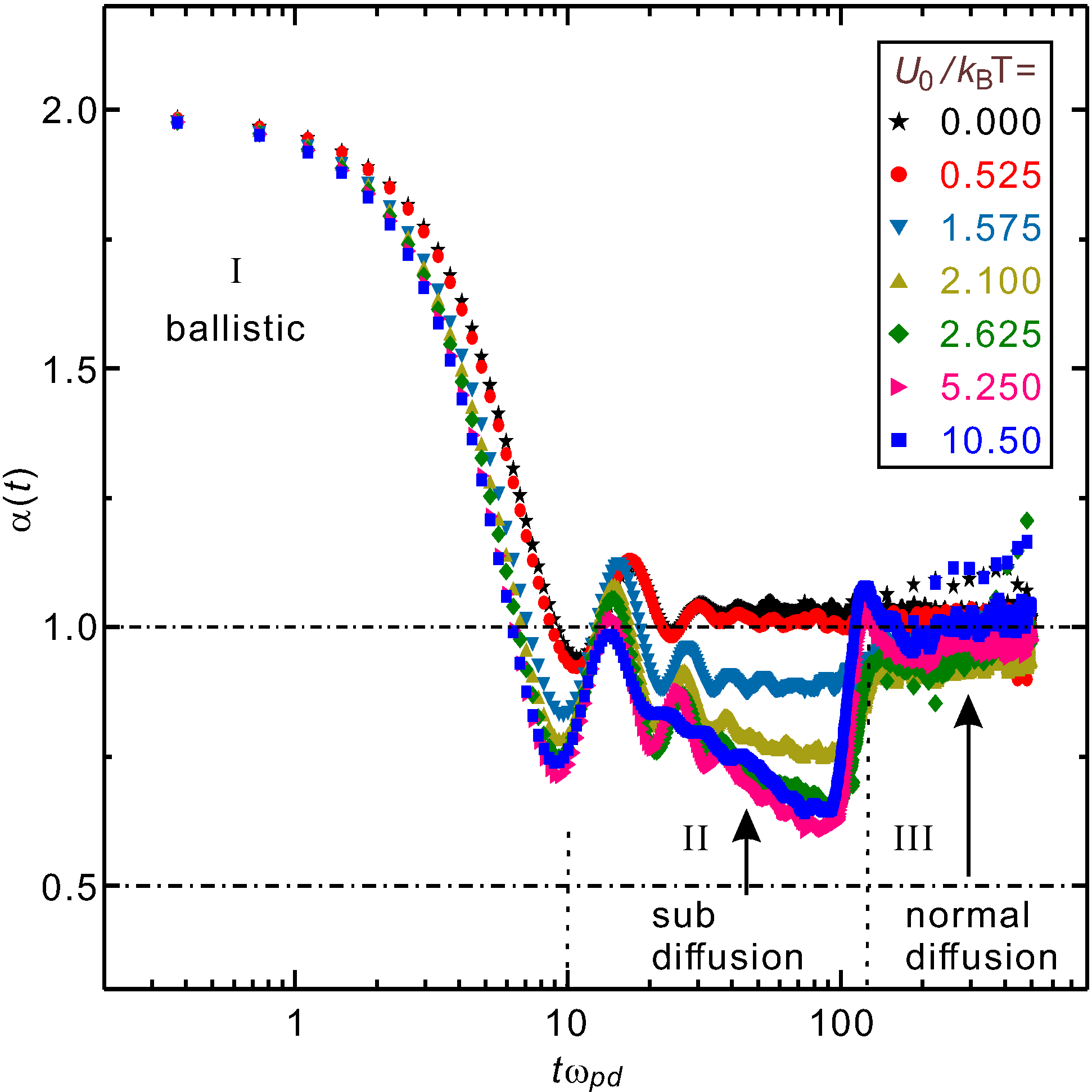}
	\caption{\label{alpha_t}(Color online) For the motion in the $y$ direction, the obtained exponent $\alpha(t)$ in the relation $MSD = 4Dt^{\alpha(t)}$ for different depths of 1DPS at different times. Here, the exponent $\alpha(t)$ in the $y$ direction are calculated using the data from Fig.~6(c). The diffusion of 2DDP motion in the $y$ direction can be divided into three processes, due to the 1DPS. Initially, the exponent decays from 2, due to the ballistic motion of dust particles, since they are still inside the cages formed by neighboring dust particles. At the intermediate time scale, the exponent of diffusion is between 0.5 and 1, corresponding to the sub-diffusion. This sub-diffusion feature is obvious, even when the depth of the 1DPS is not so large, such as $U_{0}/k_{B}T \ge 1.575$. For the long time scale, the diffusion all changes to normal diffusion ($\alpha = 1.0$) for different depths of 1DPS.
	}
\end{figure}

To quantify the properties of diffusion of 2DDP on 1DPS, we calculate the exponent $\alpha(t)$ in the above relation $MSD$ using
\begin{equation}\label{alpha}
{\alpha (t) = \frac {\partial \ln (MSD)}{\partial \ln (t)},}
\end{equation}
as described in~\cite{Hanes:2012}. Here only the diffusion in the $y$ direction, as shown in Fig.~6(c) is used to calculate $\alpha(t)$, as plotted in Fig.~7 for different substrate strength. From Fig.~7, three processes of diffusion are clearly observed, due to the presence of a 1DPS. At early times, we find superdiffusive motion with $\alpha=2$, corresponding to the initial ballistic motion~\cite{Liu:2008} of the dust particles within the cages formed by the repulsion from neighboring dust particles. For the intermediate time scales, from around $t\omega_{pd} \approx 10\sim 20$, we discover that the motion becomes increasingly subdiffusive as the substrate strength increases, with $0.5 < \alpha(t) < 1.0$. We speculate this subdiffusive motion probably is due to the substrate-induced distortion of the dynamic cage formed by the nearest-neighbor particles~\cite{Herrera-Velarde:2009}. This resembles the subdiffusive behavior found at intermediate times for 2D colloidal assemblies confined by a 1DPS~\cite{Herrera-Velarde:2007}, but it has not been observed previously in a dusty plasma before, except for our simulation here. For the time scale $t\omega_{pd} \approx 120$, as shown in Fig.~6(c) and Fig.~7, we find a transition to normal diffusion with $\alpha=1.0$ for all substrate strength, and the finally normal diffusion occurs for the long time scale. Note that the transition from the intermediate subdiffusive motion to final diffusive motion can be observed, as the legend marked in Fig.~6(c).

Note that, we have not observed any evidence for single-file diffusion with $\alpha(t) = 0.5$ at long times, despite the fact that the 2DDP forms 1D chains when the substrate is strong, which would appear to satisfy the conditions required for single-file diffusion~\cite{Lutz:2004,Herrera-Velarde:2007}. We speculate that maybe this is due to the underdamped feature of our studied dusty plasmas.

\section{IV.~Summary}

In summary, we have studied the structure and diffusion of 2D dusty plasmas on 1D periodic substrate using Langevin dynamical simulations. From the calculated averaged pair correlation function $g(y)$, structure factor $S(q)$, and the heights of the first and second peaks of both $g(y)$ and $S(q)$, we find that, as the strength of 1DPS increases from 0, the 2DDP transits from a disordered liquid state to a modulated ordered state, and finally to a modulated disordered state again with strong substrates. We speculate this reentrant melting probably results from the increasing constraint of particle motion by the substrate, which would result in a reduced effective dimensionality and simultaneously enhance the fluctuation effects.

We also present that the substrate confines the diffusion of the dust particles, and that for sufficiently strong substrates, motion along the potential wells of 1DPS dominates the diffusive behaviors. Our analysis of the MSD along the potential wells clearly indicates a subdiffusion process at intermediate times between the initial ballistic and final diffusive motions. We suggest the subdiffusive motion discovered here is probably due to the substrate-induced distortion of the dynamic cage formed by the nearest-neighbor particles.

We thank J.~Goree for helpful discussion. Work in China was supported by the National Natural Science Foundation of China under Grant No. 11505124, the 1000 Youth Talents Plan, the Six Talent Peaks project of Jiangsu Province, and startup funds from Soochow University. Work at LANL was carried out under the auspices of the NNSA of the U.S. DOE under Contract No. DE-AC52-06NA25396.

\end{document}